\documentclass[conference]{IEEEtran}
\IEEEoverridecommandlockouts
\usepackage{cite}
\usepackage{amsmath,amssymb,amsfonts}
\usepackage{algorithmic}
\usepackage{graphicx}
\usepackage{textcomp}
\usepackage{xcolor}
\usepackage{adjustbox}
\usepackage{diagbox}

\def\BibTeX{{\rm B\kern-.05em{\sc i\kern-.025em b}\kern-.08em
    T\kern-.1667em\lower.7ex\hbox{E}\kern-.125emX}}
\begin{document}

\title{Empirical Quantum Advantage Analysis of Quantum Kernel in Gene Expression Data\\
\thanks{}
}

\makeatletter
\newcommand{\linebreakand}{%
  \end{@IEEEauthorhalign}
  \hfill\mbox{}\par
  \mbox{}\hfill\begin{@IEEEauthorhalign}
}
\makeatother

\author{\IEEEauthorblockN{Arpita Ghosh}
\IEEEauthorblockA{\textit{Computer Science \& Engineering} \\
\textit{Shahjalal University of Science \& Technology}\\
Sylhet, Bangladesh \\
ghosharpita423@gmail.com}
\and
\IEEEauthorblockN{Seemanta Bhattacharjee}
\IEEEauthorblockA{\textit{Computer Science \& Engineering} \\
\textit{Shahjalal University of Science \& Technology}\\
Sylhet, Bangladesh \\
babune99@gmail.com}
\linebreakand
\IEEEauthorblockN{MD Muhtasim Fuad}
\IEEEauthorblockA{\textit{Computer Science \& Engineering} \\
\textit{Shahjalal University of Science \& Technology}\\
Sylhet, Bangladesh \\
muhtasimfuad625@gmail.com}
}

\maketitle

\begin{abstract}
The incorporation of quantum ansatz with machine learning classification models demonstrates the ability to extract patterns from data for classification tasks. However, taking advantage of the enhanced computational power of quantum machine learning necessitates dealing with various constraints. In this paper, we focus on constraints like finding suitable datasets where quantum advantage is achievable and evaluating the relevance of features chosen by classical and quantum methods. Additionally, we compare quantum and classical approaches using benchmarks and estimate the computational complexity of quantum circuits to assess real-world usability. For our experimental validation, we selected the gene expression dataset, given the critical role of genetic variations in regulating physiological behavior and disease susceptibility. Through this study, we aim to contribute to the advancement of quantum machine learning methodologies, offering valuable insights into their potential for addressing complex classification challenges in various domains.
\end{abstract}


\section{Introduction}
In the face of rapid mutations and variations in diseases, the ever-expanding volume of biological data presents unparalleled opportunities for unraveling intricate biological phenomena. Among these, gene expression analysis emerges as a cornerstone tool for comprehending the molecular mechanisms underlying diverse physiological and pathological processes. This comprehension leads to the accurate classification of cancer subtypes, which is essential for guiding individualized treatment strategies\cite{b1}. In this endeavor, the Golub et al. gene expression dataset has been serving as an instrumental resource facilitating numerous studies aimed at precisely categorizing different subtypes of leukemia based on their distinctive gene expression profiles.

However, the sheer magnitude and intricacy of gene expression data pose formidable challenges to extracting meaningful insights. Traditional classification methods often face computational limitations when tasked with discerning patterns amidst the noise and high-dimensional feature spaces inherent in gene expression data. In this context, harnessing these principles of quantum mechanics, quantum computing offers to exponentially accelerate data processing tasks, resulting in a paradigm shift in computational power and efficiency\cite{b2}.

In this paper, we conduct an experiment investigation on gene expression data, exploring and evaluating the efficacy of both classical and quantum computing approaches for solving key challenges in gene expression classification. Our approach integrates innovative methodologies in feature selection, classification algorithms, and complexity analysis to advance our understanding of biological systems. Specifically, as part of our gene expression data analysis, we utilize quantile normalization\cite{b3} to preprocess the Golub et al. dataset, ensuring uniformity and reliability of data in our subsequent analyses. After preprocessing, it's significant to filter out relevant features to simplify the model and improve computational efficiency. To conduct a comparative study, we adopt both quantum and classical approaches for this task. The paper\cite{b4} highlights the effectiveness of Lasso in identifying relevant features from high-dimensional datasets, particularly in the field of genomics where the number of features often exceeds the number of samples. So we have employed LASSO Regularization (L1)\cite{b5} for classical means of feature selection. For the quantum approach to feature selection, we utilized D-Wave's hybrid quantum-classical framework\cite{b6} leveraging D-Wave's quantum annealers\cite{b7} to formulate the Quantum Unconstrained Binary Optimization (QUBO) problem. This hybrid approach allows us to explore alternative solutions for feature selection tasks in high-dimensional datasets. 

We have classified the data using both quantum and classical kernels, utilizing the features selected by both approaches. To evaluate the performance of both the classical and quantum kernels using multiple metrics, including the F1 score, balanced accuracy, and Phase Terrain Ruggedness Index (PTRI), geometric difference\cite{b8}. Finally, we conduct a comparative analysis of the computational complexity of quantum and classical kernels to evaluate their practical feasibility in large-scale gene expression analysis.

\section{Methodology}
According to a genome-wide association study (GAWS), detecting the precise biomarker contributes vitally to diagnosing specific diseases. Gene expression data with large feature space also enables the exploitation of the exponential transformation capability of quantum embedding. The gene expression dataset (Golub et al.) generated by a proof-of-concept study for cancer subtyping task (AML vs ALL) comprises 7129 gene expression profiles for each of the 38 training and 34 test samples. As gene expression values vary in a wide range,  quantile normalization \cite{b3} is performed to make the values comparable across different samples.

Then from this large normalized data, 20 important features are extracted using \cite{b9} the L1 regularization(Lasso) method. Here are the gene accession number of the selected genes [`AB000466\_at', `D17391\_at', `D38524\_at', `D78134\_at', `HG3945-HT4215\_at', `J04990\_at', `J05158\_at', `L01664\_at', `M26602\_at', `M60047\_at', `M63904\_at', `S67156\_at', `S77094\_at', `U30828\_at', `U47011\_cds1\_at', `U63289\_at', `U66580\_at', `U70981\_at', `M15169\_at', `J00268\_s\_at'].
Simultaneously, the resultant genes from QUBO are [`D17391\_at', `HG1148-HT1148\_at', `HG4188-HT4458\_at', `L09717\_at', `M77810\_at', `S82240\_at', `U14550\_at', `U16997\_at', `U19878\_at', `U34360\_at', `U49248\_at', `U63289\_at', `X07820\_at', `X14046\_at', `X17042\_at', `U31556\_at', `M27783\_s\_at', `M63438\_s\_at', `HG3731-HT4001\_r\_at', `U84388\_at'].

To build a more robust model against the
outliers, min-max scaling is performed. After tuning the parameter, it's inferred that the range from $0$ to $\pi$ works well for this specific experiment. Support vector machine(SVM) is a widely used tool for classification tasks in supervised machine learning that applies kernel tricks when the data is not linearly separable in their original feature space. It transforms data into  higher dimensional feature space where the data is linearly separable. The efficient kernel approximation is crucial for resource optimization and performance enhancement of a model.  For this experiment, we have adopted the quantum kernel estimation method implemented by  Havlicek et al\cite{b10} where the kernel function is calculated using a quantum circuit and then it's passed to the classical SVM for drawing a decision boundary. This model is implemented using qiskit library. 

For solving certain tasks, utilizing quantum properties provides exponential speedup over the best-known classical approach, which is termed as quantum supremacy. But in near-term quantum hardware, the quantum advantage is yet to be attained for all kinds of operations. Due to the vulnerable nature of Noisy intermediate-scale quantum(NISQ) devices and the decoherence effects of qubits, the problem is critical to the size of the feature space. To verify the feasibility of the quantum approach, some heuristic metrics are applied. This paper \cite{b8} introduced a novel approach for this type of measurement named empirical quantum advantage(EQA). It’s a framework that
analyzes different performance metrics. For instance, in this experiment, the following measures have been used at
various configuration spaces: F1 score, balanced accuracy, and the phase space terrain ruggedness index (PTRI). The paper\cite{b8} also  suggests geometric difference as a framework to evaluate the kernel’s efficiency where the quantum advantage is potential. Here, the metric is applied to determine the kernel’s capability
to generate the optimized decision boundary for a classification problem.  It provides insight into which kernel outperforms the other kernel based on the shape of the decision
boundary. 

\section{Result And Discussion}

After performing all the pre-processing tasks, the dataset is split into train and test subsets with an $80:20$ ratio. The configuration space of this experiment consists of $57$ training samples. Nine sub-configuration spaces are chosen for that configuration space. The sub-configuration spaces consist of sample size $[25, 41, 57]$ and features $[2, 8, 14]$, each feature denoted by a single qubit. Then F1 score and balanced accuracy are calculated for nine configurations.

\begin{figure}[h]
   \centering
   \includegraphics[width = \linewidth]{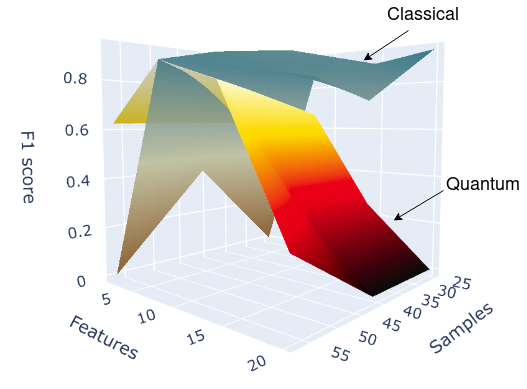}
   \caption{F1 score of data selected by lasso.}
\label{fig:fig_f1_score_lasso_surface_plot}
\end{figure}

\begin{figure}[h]
   \centering
   \includegraphics[width = \linewidth]{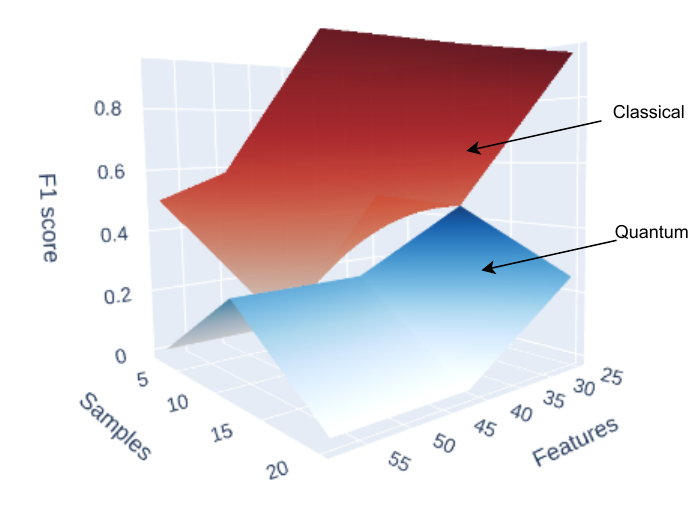}
   \caption{F1 score of data selected by QUBO.}
\label{fig:fig_f1_score_qubo_surface_plot}
\end{figure}

It is observed from the Fig. \ref{fig:fig_f1_score_lasso_surface_plot} that, for $14$ features and $25$ samples, classical kernel has the highest F1 score of $.93$ and for $8$ features and $57$ samples quantum kernel has $.85$. Classical and quantum surfaces intersect at $(8,57)$ point. And in Fig.\ref{fig:fig_f1_score_qubo_surface_plot} for $14$ features and $25$ samples, classical kernel has the highest F1 score of $.93$ and for $8$ features and $41$ samples quantum kernel has $.44$. Classical and quantum surfaces intersect at $(8,41)$ point.

\begin{figure}[h]
   \centering
   \includegraphics[width = \linewidth]{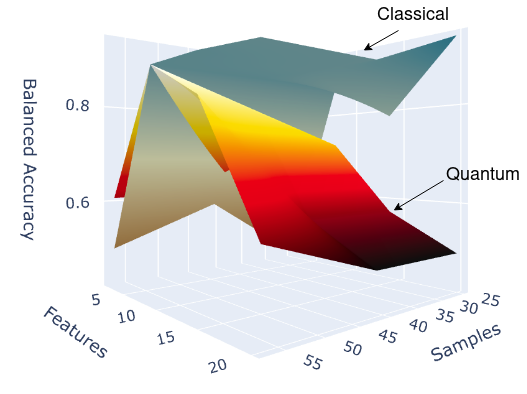}
   \caption{Balanced accuracy of data selected by lasso.}
\label{fig:fig_balanced_accuracy_lasso_surface_plot}
\end{figure}

\begin{figure}[h]
   \centering
   \includegraphics[width = \linewidth]{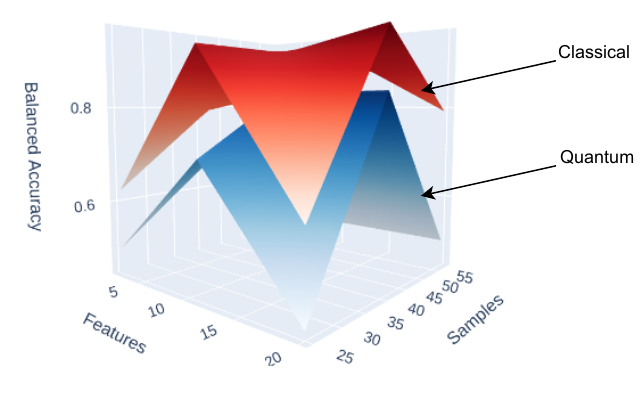}
   \caption{Balanced accuracy of data selected by QUBO.}
\label{fig:fig_balanced_accuracy_qubo_surface_plot}
\end{figure}

Similarly, from Fig. \ref{fig:fig_balanced_accuracy_lasso_surface_plot}, it is observed that for $20$ features and $25$ samples classical kernel has the highest balanced accuracy of $.93$ and for $8$ features and $57$ samples quantum kernel has balanced accuracy of $.86$. Classical and quantum surfaces intersect at $(8,57)$ point. In Fig. \ref{fig:fig_balanced_accuracy_qubo_surface_plot}, for $8$ features and $57$ samples classical kernel has the highest balanced accuracy of $.96$ and for $2$ features and $41$ samples quantum kernel has balanced accuracy of $.64$. Classical and quantum surfaces intersect at $(2,41)$ point.

\begin{figure}[h]
   \centering
   \includegraphics[width = \linewidth]{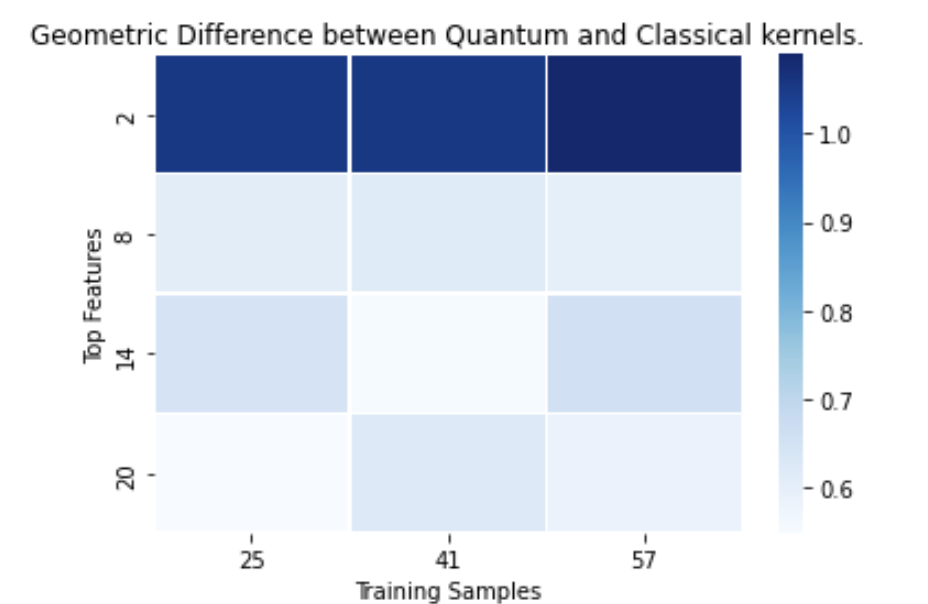}
   \caption{Heatmap for the geometric difference of both kernels.}
\label{fig:fig_geometric_dif}
\end{figure}

 For different configuration spaces, the geometric difference between classical(linear) and quantum kernel(Pauli Z feature map) is analyzed. Fig. \ref{fig:fig_geometric_dif} shows that potential quantum advantage is more likely in the configuration space with $2$ features and $57$ samples. 

 The Phase terrain ruggedness index (PTRI) is a metric that helps to identify the configuration space where quantum advantage is potential for a specific problem. The flattest region in classical landscape helps to consider quantum configuration for that problem to attain privilege over the classical. As the flattest region indicates stagnation of performance, the ruggedness of the quantum landscape helps to get insights if there is any quantum advantage possible.

 \begin{figure}[h]
   \centering
   \includegraphics[width = \linewidth]{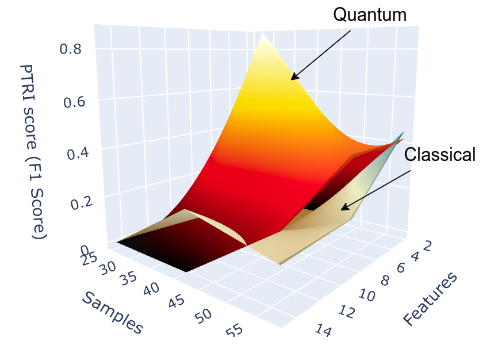}
   \caption{PTRI landscape for F1 Score of data selected by lasso.}
\label{fig:fig_ptri_f1_lasso}
\end{figure}

\begin{figure}[h]
   \centering
   \includegraphics[width = \linewidth]{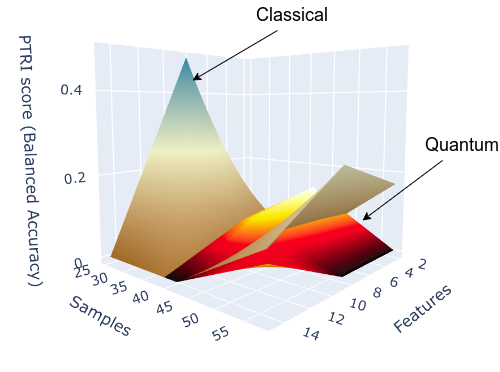}
   \caption{PTRI landscape for Balanced Accuracy Score of data selected by lasso.}
\label{fig:fig_ptri_ba_lasso}
\end{figure}

The ruggedness of the PTRI landscape for the F1 score in Fig.\ref{fig:fig_ptri_f1_lasso} indicates a point of advantage for the quantum kernel in the configuration of $(8,41)$, where the classical kernel performs poorly. 
Whereas, it can be observed from the PTRI landscape for Balanced Accuracy that classical kernel performs better in configuration of $(8,25)$ in Fig.\ref{fig:fig_ptri_ba_lasso} than the quantum counterpart. 

It is also to be considered the PTRI values resulting in calculations with different metrics should indicate which performance metric is better with accuracy for a problem in the analysis of EQA. The experiment has better performance accuracy with F1 score metric facilitating quantum advantage in the configuration of $(8,25)$. It also shows that balanced accuracy may seem a reasonable metric to proceed initially with much prediction accuracy.
But, considering PTRI to evaluate its potentiality towards quantum advantage does provide much insight into its unlikeliness of outperforming.

\section{Quantum resource estimation}
Proceeding toward solving a problem requires the consideration of the efficiency of the implementation. So it is important to analyze the complexity of an algorithm along with enhancing its performance. For this type of evaluation, resource estimation is vital in the quantum field. Quantum resource estimation is a method to determine the number of qubits, unitary gates, quantum processing unit(QPU) utilization, and other resources required for algorithmic implementation. \\

In the quantum kernel estimation  approach, the kernel is estimated using quantum unitary circuit. The quantum kernel transforms the classical state into a quantum state. Then the classical SVM is applied to draw the separating hyperplane among classes. Here, the classical data $\Vec{x} \in \Omega$
 is converted into a quantum state $|\Phi(\Vec{x})\rangle$ by applying
the unitary circuit $\mathcal{U}_{\Phi(\Vec{x})}$ , where the quantum state,
$$ |\Phi(\Vec{x})\rangle =  U_{\Phi(\Vec{x})}  H^{\otimes n} U_{\Phi(\Vec{x})}  H^{\otimes n} |0\rangle^{\otimes n}$$
$$ U_{\Phi(\Vec{x})} = \exp{\Bigg( i\sum_{S \subseteq [n]}  \phi_S(\Vec{x}) \prod_{i\in S} Z_i \Bigg)} $$
Considering maps with low-degree expansions where $|S| \leq 2$, two types of feature maps are possible.
For $d=2$, the feature map that results in,
$$U_{\phi\{k,l\}} (\Vec{x}) = \exp{(i \phi _{ \{k,l\} } (\Vec{x}) Z_k Z_l)}$$
And the circuit in Fig. \ref{fig:fig_ZZfeauturemap_qiskit} is drawn for this mapping for 3 qubits and linear entanglement. This circuit is repeated 2 times. 

\begin{figure}[h]
   \centering
   \includegraphics[width = \linewidth]{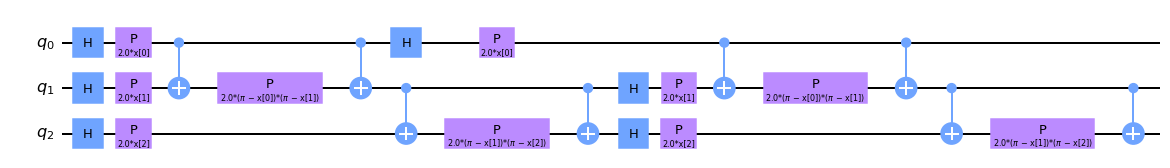}
   \caption{Circuit for ZZfeaturemap}
\label{fig:fig_ZZfeauturemap_qiskit}
\end{figure}

Using linear entanglement, for different numbers of qubits and repetitions, a depth analysis is shown in the following table, 

\begin{table}[ht]
    \centering
    \caption{Depth analysis of ZZfeaturemap with linear entanglement.}
    \begin{tabular}{|c|c|c|c|c|}

        \hline
         \diagbox[width=10em]{n}{r} & 1 & 2 & 3 & 4\\
        
        \hline
        2 & 5 & 10 & 15 & 20 \\
        \hline
        3 & 8 & 16 & 24 & 32 \\
        \hline
        4 & 11 & 19 & 27 & 35 \\
        \hline
        5 & 14 & 22 & 30 & 38 \\
        \hline 
        6 & 17 & 25 & 33 & 41 \\
        \hline
    \end{tabular}
    \label{tab:caption}
\end{table}

Here, $n$ is the number of qubits and $r$ is the number of repetitions. After analyzing the above table, it can be concluded that ZZfeaturemap has $\mathcal{O}(5 \times r)$ complexity for $n=2$, and for $n\geq 3 $ it is $\mathcal{O}(8 \times r + 3 \times (n-1)) $, both are equivalent to $\mathcal{O} (n)$. 
\\
For $d = 1$, the resulting feature map is, 
 $$ U_{\phi\{k\}} (\Vec{x}) = \exp{(i \phi _{ \{k\} } (\Vec{x}) Z_k )}$$ 
 
And the circuit in Fig. \ref{fig:fig_paulizfeaturemap} is drawn for this mapping with two qubits and two repetitions.

\begin{figure}[h]
   \centering
   \includegraphics[width = \linewidth]{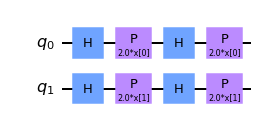}
   \caption{Circuit for Pauli Z featuremap}
\label{fig:fig_paulizfeaturemap}
\end{figure}
Observing the circuit architecture, it is evident that the circuit has $\mathcal{O} (2 \times r)$ which is equivalent to $\mathcal{O}(r)$ where $r$ is the number of repetitions.

Here is the summary of estimated gates for each feature map.

\begin{table}[h]
    \centering
    \caption{Individual gate estimation for each featuremap}
    \begin{adjustbox}{max width=\linewidth}
    \begin{tabular}{|c|c|c|c|}
        \hline
        \multicolumn{1}{|c|}{\diagbox[width=10em]{Featuremap}{Unitary gate}} & Hadamard Gate & Z gate & CNOT gate \\
        \hline
        ZZfeaturemap & $\mathcal{O}(n\times r)$ & $\mathcal{O}(r\times (2\times n -1 ))$ & $\mathcal{O}(2\times (n-1)\times r)$ \\
        \hline
        PauliZfeaturemap & $\mathcal{O}(n\times r)$ & $\mathcal{O}(n\times r)$ & - \\
        \hline
    \end{tabular}
    \end{adjustbox}
    \label{tab:caption}
\end{table}

\end{document}